\documentstyle[12pt]{article}

\begin{document}
\begin{flushright}
{CCUTH-96-05\\ IP-ASTP-04-96\\ hep-ph/9608389}
\end{flushright}
\vspace*{5mm}
\begin{center}
{\large \bf  Determination of the HQET Parameters from}\par 
{\large \bf the $B \to X_s\gamma$ Decay }\\
\vskip 1.0cm
\centerline{Hsiang-nan Li$^1$ and Hoi-Lai Yu$^2$}
\vskip 0.3cm
\centerline{$^1$Department of Physics, National Chung-Cheng University,}\par
\centerline{Chia-Yi, Taiwan, R.O.C.}
\vskip 0.3cm
\centerline{$^2$ Institute of Physics, Academia Sinica, Taipei, Taiwan, 
R.O.C.}\par

\vspace*{0.75cm}
{\it \today}\\
\vspace*{0.5cm}

\vspace*{1.0cm}
{\bf Abstract}
\vskip 0.3cm
\end{center}
\baselineskip=2.0\baselineskip

We combine the resummations for radiative corrections and for the 
heavy quark expansion to study the inclusive radiative decay 
$B \to X_s\gamma$. The infrared renormalon ambiguity is also taken into
account. Including both theoretical and 
experimental uncertainties, we determine the allowed domain for the HQET 
parameters ${\bar \Lambda}$ and $\lambda_1$ centered at
${\bar \Lambda}=0.65$ GeV and $\lambda_1=-0.71$ GeV$^2$.

\newpage

With the progress of the heavy quark effective theory (HQET), we have
gained better insight to the dynamics of heavy systems. For example, the
inclusive semileptonic $B$-meson decay rates can be expanded 
in powers of $1/m_b$. To $O(1/m_b^2)$ 
({\it ie} within an accuracy of about 5\%), three parameters
${\bar \Lambda} = m_B-m_b$, $\lambda_1=\langle (i D)^2\rangle$ 
and $\lambda_2 =\langle\sigma \cdot G\rangle$ are relevant. Except for
$\lambda_2$ which can be obtained directly from the $B^*$-$B$ mass 
splitting, ${\bar\Lambda}$ and $\lambda_1$ are poorly determined.
Recently, we have formulated the perturbative QCD (PQCD) approach to 
the inclusive semileptonic $B$-meson decays, which 
combines the resummation technique and 
the HQET based operator product expansion 
\cite{LY}. In this approach the charged lepton spectrum is expressed 
as the convolution of a hard subprocess with a jet function and a 
universal $B$-meson distribution function. 

In this letter we shall determine the parameters $\bar\Lambda$ and 
$\lambda_1$, which are equivalent to the first and second 
moments of the $B$-meson distribution function, respectively, from
the photon energy specrum of the radiative 
decay $B \to X_s\gamma$ \cite{A}. It should be emphasized that the
relation ${\bar \Lambda} = m_B-m_b$, and thus the extraction of the pole
mass $m_b$ of the $b$ quark, suffer the infrared (IR) renormalon ambiguity
of power $\Lambda_{\rm QCD}/m_b$ \cite{BSUV}. Hence, we 
regard ${\bar\Lambda}$ as being related to the first moment of the 
distribution function, instead of to the $b$ quark mass. This viewpoint
is satisfactory enough in the sense that the moments extracted here
can be consistently employed to make predictions for other processes
because of the universality of distribution functions. 
This makes possible a model-independent determination of the 
Cabibbo-Koboyashi-Maskawa (CKM) matrix elements $|V_{cb}|$ and $|V_{ub}|$ 
from the inclusive semileptonic decays. 

On the other
hand, the ambiguity of $O(\Lambda_{\rm QCD}/m_b)$ in the definition of 
the pole mass turns out to be cancelled by the ambiguity contained in
loop corrections \cite{BSUV}, when one evaluates the total decay widths
of heavy mesons. In the factorization theorem formulated at the meson 
level, we adopt the $B$-meson mass $m_B$ and thus
avoid the ambiguity in $m_b$. It is then found that the IR renormalon
ambiguity appearing in the perturbative resummation starts at 
$O(\Lambda_{\rm QCD}^2/m_b^2)$, consistent with the conclusion in 
\cite{BSUV}. This $O(\Lambda_{\rm QCD}^2/m_b^2)$ ambiguity will be taken
into account as a source of theoretical uncertainties.
After including the uncertainty from 
the experimental data, we determine
${\bar \Lambda}=0.65^{+0.42}_{-0.30}$ GeV and 
$\lambda_1=-0.71^{+0.70}_{-1.16}$ GeV$^2$.

The first two moments of the photon energy spectrum of the decay 
$B\to X_s\gamma$ have been computed using the expansions in $1/m_b$ and 
in $\alpha_s$ \cite{KL}. Without considering the errors of data, the
parameter $\bar\Lambda\approx 450$ MeV was extracted. In our approach
the PQCD expansion in $\alpha_s$ is resummed up to next-to-leading
logarithms, and the nonperturbative heavy quark expansion in $1/m_b$ is 
resummed into the $B$-meson distribution function. 
Furthermore, both the IR renormalon ambiguity as a source of theoretical 
uncertainties and the errors of data as a source of experimental 
uncertainties are included. Therefore, our analysis is more complete.
In \cite{KLP} the contributions to the $B\to X_s\gamma$ decay from the 
$b\to sg$ transition through the $s\to \gamma$ and $g\to\gamma$ 
fragmentation functions were studied, and found to be negligible. This
observation hints that we concentrate only on the $b\to s\gamma$
transition. Single logarithms in the fragmentation functions were summed 
using the renormalization-group (RG) method. In this work we employ the 
more sophiscated resummation technique to organize the double logarithms
involved in the $b\to s\gamma$ decay.

The one-loop constraint on $\bar \Lambda$ and $\lambda_1$, 
$\bar\Lambda > [0.32-0.07(\lambda_1/0.1 {\rm GeV}^2)]$ GeV,
has been obtained from the first moment of the invariant mass spectrum of 
the $B\to X_cl\nu$ decay using the same expansion in both $1/m_b$ and 
$\alpha_s$ \cite{FLS}. This constraint, however, defines 
only an open domain. QCD sum rules are an alternative
approach to the determination of the nonperturbative parameters, which give 
$\lambda_1=-0.6\pm 0.1$ GeV$^2$ \cite{BB}. 
It is also worthwhile to mention the lattice extractions
of ${\bar m}_b(m_b)=4.17\pm 0.06$ GeV \cite{CGMS} and
${\bar m}_b(m_b)=4.0\pm 0.1$ GeV \cite{D}, which are close to the lower
bound of our predictions. It is obvious that our 
predictions are not only concrete, but consistent with the conclusions
in the literature. 

The effective Hamiltonian for the process $b \to s \gamma$ written 
in terms of dimension-6 operators is 
\begin{equation}
H_{\rm eff}(b \to s\gamma)=-\frac{G_F}{\sqrt{2}} \lambda_t 
\sum^8_{j=1} C_j(\mu){\cal O} _j(\mu)\;,
\end{equation}
with $G_F$ the Fermi coupling constant, $C_j(\mu)$ the Wilson 
coefficients evaluated at the scale $\mu$, and 
$\lambda_t=V_{tb} V^*_{ts}$ the product of the CKM matrix elements. 
The definition of the operators $ {\cal O}_j$ is referred to \cite{AG}.
Here we show only the relevant operator
\begin{equation}
{\cal O}_7=\frac{e}{16\pi^2} {\bar s_{\alpha}} \sigma^{\mu \nu} (m_b(\mu) R
+m_s(\mu)L) b_{\alpha} F_{\mu \nu}\;, 
\end{equation}
with $R=(1+\gamma_5)/2$ and $L=(1-\gamma_5)/2$. Since the mixing between 
the operators ${\cal O}_1, ..., {\cal O}_6$ and
the operators ${\cal O}_7$ and ${\cal O}_8$ vanishes at one-loop level under 
the infinite renormalization, two-loop calculations of $O(e g^2_s)$ 
and $O( g^3_s)$ are necessary in order to obtain $C_7(\mu)$ and 
$C_8(\mu)$ in the leading logarithmic approximation, where $e$ and 
$g_s$ are the electromagnetic and strong couplings, respectively. 
Recently, the $O(\alpha_s)$ 
corrections to $C_7$ and $C_8$ have been computed in \cite{AY}. 
For the $b\to s\gamma$ transition, the contributions from all the
operatorrs can be included by simply employing the effective 
Wilson coefficient $C_7^{\rm eff}=C_7+Q_dC_5+3Q_dC_6$ \cite{AG},
$Q_d$ being the charge of the $d$ quark.

The analysis in \cite{AG} involves only lowest-order PQCD corrections to 
the spectator model, and thus addresses nothing on the nonperturbative 
parameters. However, the explicit $O(\alpha_s)$ expressions for the decay 
width of $b\to s\gamma$ help to understand our formalism. The infrared 
single pole appearing in the one-loop calculation is absorbed into the 
distribution function, the double logarithms $\ln^2 (m_s/m_b)$ with $m_s$ 
the $s$ quark mass, are absorbed into the jet function, and the 
single logarithms $\ln(\mu/m_b)$ are absorbed into the hard part \cite{LY}.
Employing the resummation technique and the RG method,
these large corrections are grouped into a Sudakov factor. 
This systemmatic summation of large logarithms has not been achieved
in the literature.
The distribution function can be regarded as the consequence of the 
resummation of the heavy quark expansion \cite{N}.

The factorization formula is then written as 
\begin{equation}
\frac{1}{\Gamma^{(0)}} \frac{d \Gamma}{d E_\gamma }=m_B
\int^1_x dz \int b db f(z, b)J(z,x,m_B,b) H(x)\exp[-S(m_B,b)]\;,
\label{dr1}
\end{equation}
with the tree-level decay rate $\Gamma^{(0)}$, the jet function $J$, the 
hard part $H$, and the Sudakov exponent $S$ given by
\begin{eqnarray}
& &\Gamma^{(0)}=\frac{m_B^5}{32\pi^4}|C_7^{\rm eff}(m_B)G_F\lambda_t|^2
\alpha_{\rm em}\;,\\
& &J=J_0(\sqrt{z-x}m_Bb)\;,\;\;\;\; H=x^2\;,\\
& &S=2\int_{1/b}^{m_B}\frac{dp}{p}\int_{1/b}^{p}\frac{d\mu}{\mu}
A(\alpha_s(\mu))-\frac{5}{3}\int_{1/b}^{m_B}\frac{d\mu}{\mu}
\frac{\alpha_s(\mu)}{\pi}\;, \label{a}\\
& &A={\cal C}_F\frac{\alpha_s}{\pi}
+\left[\frac{67}{9}-\frac{\pi^2}{3}-\frac{10}{27}n_f\right]
\left(\frac{\alpha_s}{\pi}\right)^2\;,
\end{eqnarray}
with ${\cal C}_F=4/3$ the color factor. The variable $x$ is defined by 
$x=2 E_\gamma/m_B$, $E_\gamma$ being the photon energy. 
Note that the second term of $A$ is a two-loop result \cite{L}. 
For consistency, the two-loop expression of $\alpha_s(\mu)$ is
inserted into the integral of $S$. Note that
the $B$-meson mass $m_B$, instead of the $b$-quark mass $m_b$, appears
in the expression of $\Gamma^{(0)}$ as stated before. The choice
of the scale $m_B$ for $C_7^{\rm eff}$, the same as the upper bound of the 
evolution of the Sudakov factor, follows the three-scale factorization
theorem developed recently \cite{CL}. 

The argument $z$ of the function $f(z,b)$ is
the momentum fraction, and $b$ is the conjugate variable of the transverse 
momentum carried by the $b$ quark.
Including these transverse degrees of freedom, we 
set $m_s$ to zero, and let $1/b$ play the role of an infrared cutoff
in the resummation for the jet function as shown in Eq.~(\ref{a}).
Since the intrinsic $b$ dependence (the perturbative
$b$ dependence has been collected into the exponent $S$) 
is not known yet, we take the ansatz $f(z,b)=f(z)\exp(-\Sigma(b))$,
which leads to 
$\Sigma\to 0$ as $b\to 0$ according to the definition $f(z,b=0)\equiv
f(z)$. It is also natrual to assume $\Sigma >0$ for all $b$
from the viewpoint that the $b$ quark is bounded inside the $B$ meson. 
Hence, the intrinsic $b$ depedence provides further suppression.
The nonperturbative function $f(z)$, 
identified as the $B$-meson distribution function, can be
expressed as the matrix element of the $b$ quark fields, whose first
three moments are \cite{N}
\begin{eqnarray}
& &\int_0^1f(z)dz=1\;,\nonumber \\
& &\int_0^1f(z)(1-z)dz=\frac{\bar\Lambda}{m_B}+
O(\Lambda_{\rm QCD}^2/m_B^2)\;,\nonumber \\
& &\int_0^1f(z)(1-z)^2dz=\frac{1}{m_B^2}\left({\bar\Lambda}^2-
\frac{\lambda_1}{3}\right)+O(\Lambda_{\rm QCD}^3/m_B^3)\;.
\label{mo}
\end{eqnarray}

Though the exponent $\Sigma$ is unknown, we can, however, extract its
leading behavior by means of the IR renormalon analysis. Note
that the perturbative Sudakov factor $e^{-S}$ in Eq.~(\ref{dr1}) becomes
unreliable as the transverse distance $b$ approaches $1/\Lambda_{\rm QCD}$.
Near this end point, $\alpha_s(1/b)$ diverges, and IR renormalon
contributions are significant. We reexpress the RG
result of the evolution of the distribution function, which is contained
in the second term of $S$, as 
\begin{equation}
W=\exp\left[4\pi{\cal C}_F\int\frac{d^{4}l}
{(2\pi)^{4}}\frac{v_\mu v_\nu}{(v\cdot l)^2}2\pi\delta(l^2)
\alpha_s(l_T^2)e^{i{\bf l}_T\cdot {\bf b}}N^{\mu\nu}\right]\;.
\label{p1}
\end{equation}
The loop integral corresponds to the correction from a real soft gluon 
attaching the two valence $b$ quarks, whose propagators have been 
replaced by the eikonal lines in the direction $v=(1,1,{\bf 0})$ \cite{L}.
The tensor $N^{\mu\nu}=g^{\mu\nu}-(n^\mu l^\nu+l^\mu n^\nu)/(n\cdot l)
+n^2l^\mu l^\nu/(n\cdot l)^2$ comes from the gluon propagator in 
axial gauge $n\cdot A=0$. We have set the argument of the running 
$\alpha_s$ to $l_T^2$, which is conjugate to the scale $b$ of the
distribution function.

Substituting the identity $\alpha_s(l_T^2)=\pi\int_0^{\infty}d\sigma 
\exp[-2\sigma \beta_1\ln(l_T/\Lambda_{\rm QCD})]$, $\beta_1=(33-2n_f)/12$,
into Eq.~(\ref{p1}), and performing the loop integral for
$n\propto (-1,1,{\bf 0})$ \cite{L}, we obtain
\begin{equation}
W=\exp\left[{\cal C}_F\int_0^{\infty}d\sigma
\left(\frac{b\Lambda_{\rm QCD}}{2}\right)^{2\sigma\beta_1}
\frac{\Gamma(-\sigma\beta_1)}{\Gamma(1+\sigma\beta_1)}\right]\;.
\end{equation}
It is easy to observe that the pole of $\Gamma(-\sigma\beta_1)$ 
at $\sigma\to 0$ gives the perturbative anomalous dimension of the 
distribution function appearing in Eq.~(\ref{a}). The extra poles at 
$\sigma\to 1/\beta_1$, $2/\beta_1$,..., then correspond to the IR 
renormalons, giving corrections of powers $b^2$, $b^4$,..., respectively. 
These renormalons
generate singularities, which must be compensated by the nonperturbative
power correctons in order to have a well-defined perturbative expansion. 
Note that the renormalon ambiguity starts with the power 
$(b\Lambda_{\rm QCD})^2$, instead of $b\Lambda_{\rm QCD}$ \cite{BSUV}, 
because the function $W$ does not include the self-energy corrections
to the $b$ quark, which vanish under the eikonal approximation.

Using the ''minimal" ansatz \cite{KS}, {\it ie} picking up only the 
leading (left-most) renormalon contribution, we parametrize the exponent 
$\Sigma(b)$ by $\Sigma(b)=c'm_B^2b^2$, corresponding to the fact that the 
power corrections start at $O(b^2)$. Certainly, other parametrizations 
consistent with this fact are equally good. The IR renormalon ambiguity 
in other terms of $S$ can be absorbed into $\Sigma(b)$. Combined with the 
Sudakov exponent, which is approximated by $S\approx 0.025 m_B^2b^2$, 
Eq.~(\ref{dr1}) gives the differential branching ratio of the decay 
$B\to X_s\gamma$,
\begin{equation}
\frac{d BR}{d E_\gamma }=r\frac{x^2}{c}\int^1_x dz f(z)e^{-(z-x)/c}\;,
\label{dr3}
\end{equation}
with $c=4c'+0.1$ and $r=2\Gamma^{(0)}\tau_B/m_B$. 
By varying the parameter $c$ under the constraint $c > 0.1$, the theoretical
uncertainty arising from our formalism is taken into account. 
Assuming the values 
$m_B=5.279$ GeV, $G_F=1.16639\times 10^{-5}$ GeV$^{-4}$ \cite{M},
$C_7^{\rm eff}(m_B)=-0.306\pm 0.050$, $|\lambda_t|=0.040\pm 0.004$,
$\alpha_{\rm em}=1/130$ \cite{AG}, and the $B$-meson lifetime
$\tau_B=1.60\pm 0.03$ ps \cite{K},
we have $r=1.90\times 10^{-4}$ GeV$^{-1}$ with about 50\% uncertainty.
We find that the allowed values of
$c$ run from 0.1 to 0.2, {\it ie} $c'$ associated with the
nonperturbative exponent is a small number. It indicates that perturbative
corrections are more important than nonperturbative corrections,
consistent with the requirement from heavy quark symmetry.

Since two parameters are to be determined, the function $f(z)$ is modeled 
by the two-parameter form proposed in \cite{LY},
\begin{equation}
f(z)=N\frac{z(1-z)^2}{[(z-a)^2+\epsilon z]^2}\;,
\end{equation}
where $N$ is the normalization constant, and the parameters $a$ and 
$\epsilon$ will be adjusted to fit the CLEO data within errors. 
Note that the specific form of $f(z)$ is not essential. Any two-parameter
wave function serves the same purpose. $\bar\Lambda$ and $\lambda_1$ are 
then obtained from the moments of $f(z)$ using Eq.~(\ref{mo}). 
The numerical analysis proceeds in the following way: We choose
the central value $r=1.90\times 10^{-4}$ GeV$^{-1}$ first, and 
start with a value of $a\le 1$. For an arbitrary value of $\epsilon$,
we vary $c$ under the constraint $c>0.1$. If there exists a finite
range of $c$ such that the predictions from Eq.~(\ref{dr3}) fall into 
the error bars of the CLEO data 
\cite{A}, this $\epsilon$ is acceptable. Repeating this 
procedure for different $\epsilon$, an allowed range 
of $\epsilon$ is found. By changing $a$, we obtain the 
allowed domain for $a$ and $\epsilon$, from which the 
corresponding domain of $\bar \Lambda$ and $\lambda_1$ is determined.

Results are shown in Fig.~1, in which the solid curves correspond to
$a=0.91$, 0.92,..., and 1.00 from right to left,
and a point on each curve corresponds to a value of $\epsilon$.
For $a\le 0.90$, no acceptable $\epsilon$ exists. These curves form the 
corresponding allowed
domain of $\bar\Lambda$ and $\lambda_1$. Including the errors of $r$, 
the domain enlarges by 20\%. If there is no the constraint $c>0.1$,
the upper bound of $\bar\Lambda$ will increase 30\%.
The dashed curves are quoted 
from \cite{FLS}, with the left one and the right one obtained from the 
first and the second moments of the invariant mass spectrum of 
the $B\to X_cl\nu$ decay, respectively. The space above the dashed
curves is the allowed domain in \cite{FLS}. The third information from 
the ratio of 
partial widths $R_\tau=\Gamma(B\to X_c\tau\nu)/\Gamma(B\to X_ce\nu)$
introduces a constraint from the top. However, this constraint is not 
yet convincing \cite{FLS}, and still leaves the allowed domain an open one.
Though the overlap of our results with those in \cite{FLS} is not very
large, it is reasonable to claim that they are consistent with each other,
because the theoretical uncertainties of the approach in \cite{FLS}
was not estimated.

We take the middle of the $a=0.95$ curve as the central 
values, which lead to ${\bar \Lambda}=0.65^{+0.42}_{-0.30}$ GeV and 
$\lambda_1=-0.71^{+0.70}_{-1.16}$ GeV$^2$. 
The bounds of these extractions are indeed very large. However, we 
emphasize that they are the consequence of the inclusion of as more as 
possible theoretical and experimental uncertainties into our formalism. 
These bounds will definitely be reduced when, for example, more accurate 
data are available. If the errors of the data become half of current ones, 
we shall obtain ${\bar \Lambda}=0.65^{+0.34}_{-0.13}$ GeV and 
$\lambda_1=-0.71^{+0.57}_{-1.06}$ GeV$^2$. 
As a simple estimation, we extract from $\bar\Lambda$ the $b$-quark mass
$m_b=4.63^{-0.42}_{+0.30}$ GeV, which certainly suffers the IR renormalon
ambiguity. 
Employing the central 
values, we determine the $B$-meson distribution function
\begin{equation}
f(z)=\frac{0.02647z(1-z)^2}{[(z-0.95)^2+0.0034 z]^2}\;,
\label{fl}
\end{equation}
which can be used to determine the inclusive semileptonic and 
nonleptonic decay spectra  in the future.

To highlight the resummation effect, we present the predictions for the 
photon energy spectra of the
$B\to X_s\gamma$ decay derived from the nonperturbative HQET distribution 
function alone, and from our formula including the perturbative 
resummation in Eq.~(\ref{dr3}). The expression of the former is simply
$d BR/d E_\gamma=rx^2f(x)$. Substituting $r=1.90$ and Eq.~(\ref{fl}) into 
the above expression and into Eq.~(\ref{dr3}) with $c=0.15$, we obtain
the spectra shown in Fig.~2. The CLEO data are also displayed. It is obvious
that the spectrum from $f(x)$ has a sharp peak near the high end
of $E_\gamma$, which
satisfies the HQET, but is in conflict with the data. The 
predictions match the data only after the suppression effect is included, 
which possess a softer profile. Note that a naive fitting without the
suppression effect needs a board distribution function, leading
to a value of $m_b$ as small as 3.0 GeV.

The branching ratio can be evaluated simply by integrating Eq.~(\ref{dr3})
over $E_\gamma$. We obtain $BR=2.80^{+0.14}_{-0.50}\times 10^{-4}$, where
the central value corresponds to the solid curve in Fig.~2 ($c=0.15$), and 
the upper bound and lower bound to $c=0.13$ and  $c=0.25$,
respectively. Our prediction is close to the standard-model estimation
$(2.8\pm 0.8)\times 10^{-4}$ \cite{BMMP}, and to the CLEO data
$(2.75\pm 0.67)\times 10^{-4}$ from the $B$-reconstruction analysis
\cite{A}.

This work was supported by the National Science Council of the Republic of
China under Grand Nos. NSC-85-2112-M-194-009 and NSC-85-2112-001-021.

\newpage
\begin{center}
{\large\bf Figure Captions}
\end{center}

\vskip 1.0cm
\noindent
{\bf Fig.1}: The allowed domain for $\bar\Lambda$ and $\lambda_1$.
The dashed curves are the constraints quoted from \cite{FLS}.

\noindent
{\bf Fig.2}: The photon energy spectra derived from the $B$-meson
 distribution 
function alone (dashed curve), and from the inclusion of the suppression 
effect (solid line). The CLEO data are also shown \cite{A}.

\end{document}